\documentstyle[12pt,aaspp4,psfig]{article} % local pp version
\begin{document}                           
 
\title{Recent Outbursts from the Transient X-Ray Pulsar Cep X--4 (GS 2138+56)}
\author{Colleen~A.~Wilson, Mark~H.~Finger\altaffilmark{1}, D.~Matthew Scott\altaffilmark{1}}
\affil{\footnotesize ES 84 Space Sciences Laboratory, NASA/Marshall
Space Flight Center, Huntsville, AL 35812;
colleen.wilson@msfc.nasa.gov, finger@gibson.msfc.nasa.gov,
scott@gibson.msfc.nasa.gov}
\altaffiltext{1}{Universities Space Research Association}

\begin{abstract}

We report on X-ray observations of the 66 s period transient X-ray pulsar Cep 
X--4 (GS 2138+56) with the Burst and Transient Source Experiment (BATSE) on the 
{\em Compton Gamma-Ray Observatory (CGRO)} and with the Rossi X-ray Timing
Explorer ({\em RXTE}). Two outbursts from Cep X--4 were observed with BATSE in 
1993 June-July and 1997 July. Pulse frequencies of $\nu = 15.0941 \pm 0.0002$ mHz on 
1993 June 25 (MJD 49,163) and $\nu = 15.0882 \pm 0.0002$ mHz 
on 1997 July 12 (MJD 50,641) were each measured from 2 day spans of BATSE data 
near each outburst's peak. Cep X--4 showed an average spin down rate of 
$\dot \nu = (-4.14 \pm 0.08) \times 10^{-14}$ Hz s$^{-1}$ between the 1993 and 1997 
outbursts. After BATSE could no longer detect Cep X--4, public observations 
were performed on 1997 July 18 \& 25 with the Proportional Counter Array (PCA)
on {\em RXTE}. A pulse frequency of $\nu = 15.088 \pm 0.004$ mHz was measured 
from observations on 1997 July 18 (MJD 50,647). Significant aperiodic noise, 
with an rms variance of $\sim$18\% in the frequency range 0.01-1.0 Hz was 
observed on both days. Energy and intensity dependent pulse shape variations 
were also seen in these data. Recently published optical observations 
associate Cep X--4 with a Be companion star. If all 4 outbursts observed from 
Cep X--4 are assumed to occur at the same orbital phase, we find that the
orbital period is between 23 days and 147.3 days.
\end{abstract}


\section{Introduction} 

In the last 25 years more than 40 accretion-powered X-ray pulsars have been
discovered.  At least half of these are transient, of which 12 have known Be star
companions.  Neutron stars with Be companions accrete material from the slow,
dense, stellar outflow thought to be confined to the equatorial plane of the Be
star. Cep X--4 has recently been associated with a Be star of magnitude V=14.2 
(\cite{Roche97}; \cite{Argyle97}) that lies within the {\em ROSAT} error circle
(\cite{Schulz95}). This association is based upon positional coincidence, 
strong H-$\alpha$ and H-$\beta$ emission lines (typical of Be/X-ray binaries),
and measurement of the optical absorption column density which is in good agreement 
with {\em ROSAT} measurements. The companion is most likely a B1V-B2Ve star at 
a distance of 3.8$\pm$0.6 kpc (\cite{Bonnet98}).

Cep X--4 was discovered with {\em OSO-7} in 1972 June-July, but no
pulsations were detected (\cite{Ulmer73}). The {\em Ginga} All-Sky
Monitor first detected a new transient source, GS 2138+56, on 1988 March
19 at a 50 mcrab level. GS 2138+56 was later identified with Cep X--4 because 
it fell in the 0.4\arcdeg\ error box and had a similar spectrum. Pulsations at
a frequency of 15.09457 $\pm$ 0.00002 mHz (MJD 47,263.5) were detected during a month 
long outburst which peaked at about 100 mCrab (1-20 keV) in early April 1988.
The source apparently did not appear again until 1993 June when it was 
detected with {\em ROSAT} (\cite{Schulz95}) and the Burst and Transient Source
Experiment (BATSE) on the {\em Compton Gamma-Ray Observatory (CGRO)}. Near the
outburst peak,
pulsations at a frequency of 15.0941 $\pm$ 0.0002 mHz (MJD 49,163.0) were detected 
with BATSE.  The outburst lasted about two weeks and had a peak pulsed flux of
15-20 mCrab (20-50 keV). In 1997 July-August, BATSE and the All-Sky Monitor (ASM) on 
the {\em Rossi X-ray Timing Explorer (RXTE)} observed
an outburst from Cep X--4. BATSE detected pulsations at a frequency of
15.0882 $\pm$ 0.0002 mHz (MJD 50,641) near the outburst peak. This outburst lasted about
2 weeks and peaked at a pulsed flux of about 10-15 mCrab (20-50 keV). The ASM
observed a longer (~4 week) outburst with a peak flux of about 40 mCrab (2-12
keV). The best
fit average spin-down rate is $\dot \nu = (-4.1 \pm 0.4 ) \times 10^{-15}$ 
Hz s$^{-1}$ between the 1988 and 1993 outbursts. Between the 1993 and 1997 
outbursts, the average spin-down rate, $\dot \nu = (-4.14 \pm 0.08) \times 
10^{-14}$ Hz s$^{-1}$, is 10 times larger than the spin-down rate between the first
two outbursts. Apparent spin-up observed during the first two outbursts and 
spin-down in the recent outburst may be intrinsic or due to binary orbital motion.

In this paper we present observations of two outbursts from Cep X--4 in
1993 and 1997. Our observations with BATSE include histories of pulse frequency, pulsed
flux, and average 20-50 keV pulse profiles. We obtained public 
{\em RXTE} Proportional Counter Array (PCA) data from 1997 July 18 \& 25, after BATSE stopped detecting Cep
X--4. We present pulse frequency measurements, pulse profiles, and power spectra
from these data.  

\section{Observations and Analyses}

\subsection{Instruments}

Three different instruments spanning energies from 2-1800 keV were used to
study Cep X--4: BATSE on  {\em CGRO}, and the PCA and ASM both on {\em RXTE}.
BATSE consists of eight identical uncollimated detector modules positioned on 
the corners of the {\em CGRO} spacecraft such that the normal vectors of the 
detectors are perpendicular to the faces of a regular octahedron, providing 
all-sky coverage.  The BATSE data presented here are taken with the large-area
detectors (LADs), which are NaI(Tl) scintillation crystals with a geometric 
area of 2025 cm$^2$ and a thickness of 1.27 cm.  The LADs are sensitive to 
photons from 20 to 1800 keV.  Two BATSE data types were used in this analysis,
the CONT (2.048 s, 16 energy channel) data and the DISCLA (1.024 s, 4 energy 
channel) data.  A more complete description of the instrument and data types 
can be found in \cite{Fishman89}.  

The PCA on {\em RXTE} consists of five 
proportional counters sensitive to photons from 2-60 keV and has a total 
collecting area of 6500 cm$^2$. The PCA is a collimated instrument with an 
approximately circular field-of-view with a FWHM of about 1\arcdeg\ 
(\cite{Jahoda96}). Three PCA data types were used in this analysis, Standard~1
(125 ms, no energy resolution) data, Good Xenon data (1 $\mu$s, 256 energy 
channel) data, and Standard~2 (16 s, 128 channel) data.
The {\em RXTE} ASM consists of three wide-angle
shadow cameras equipped with Xenon proportional counters with a total
collecting area of 90 cm$^2$.  The ASM provides 90 second images of most of
the sky every 96 minutes in three energy channels from 2-12 keV 
(\cite{Levine96}). For this analysis we used single day averaged data from the
full 2-12 keV band. 

\subsection{Techniques \label{sec:tech}}

Histories of pulse frequency and pulsed flux (20-50 keV) were determined on 
2-day intervals of BATSE DISCLA data. Pulse frequencies were estimated from
fits to data at a range of trial frequency offsets from a pulse phase model. The
pulse phase model used a constant frequency, $\nu_0$, obtained from daily power
spectra of the BATSE DISCLA data. Each 2-day interval of data were fit with a 
background model plus a 6 harmonic Fourier expansion in the pulse phase model
to generate a pulse profile at the model frequency. The background was modeled
as a quadratic spline, with segments every 300 s and with value and slope
continuous between adjacent segments. (See \cite{Bildsten97} for a detailed 
description of pulsed flux and pulse frequency estimation techniques.) 
For each 2-day span a range of trial frequency offsets, $\pm$ 3.5 cycles 
day$^{-1}$, from the model frequency was searched. For each trial frequency
offset, $\Delta \nu_m$, the pulse profile at the model frequency was shifted in
phase by multiplying the Fourier harmonic coefficients by a phase factor 
$e^{i 2 \pi k \Delta \phi_{m}}$, where $k$ is the harmonic number.  The phase 
offset, $\Delta \phi_{m}$, is given by $\Delta \phi_{m} = \Delta \nu_m 
(t - t_o)$ where $t$ is a time within to a 2-day interval and $t_o$ is 
an epoch near the center of the 2-day interval. This method is equivalent to refitting the background model plus
Fourier expansion for each frequency $\nu_0 + \Delta \nu_{m}$. The best fit 
frequency for each 2-day interval was determined by comparing pulse profiles
for different frequency offsets using the Z$^2_6$ test (\cite{Buccheri83}) 
which measured the significance of the first 6 Fourier amplitudes.  

The root-mean-square (RMS) pulsed flux was estimated from the best fit pulse 
profile as 
\begin{equation}
F_{\rm RMS} = \left[ \int^1_0 (F(\phi) - \bar F )^2 d\phi \right]^{1/2}
\end{equation}
where $F(\phi)$ is the pulse profile at phase $\phi$, $0 \leq \phi \leq 1$, and
$\bar F = \int^1_0 F(\phi)d\phi$ is the average flux. The pulsed fluxes were
determined assuming an exponential energy spectrum,
\begin{equation}
f(E) = A \exp(-\frac{E}{E_{\rm fold}})\label{eqn:spec}
\end{equation}
with a normalization $A$ and e-folding energy $E_{\rm fold}$.                                       

Pulse frequencies were estimated from {\em RXTE} PCA data using a similar method.
Since both data sets were quite short, 2000-2500 s, a model consisting of a
constant background plus an $n$ harmonic Fourier expansion in the pulsed phase
model was fit to the data. 
In both observations, The pulsed phase model used a constant
frequency, based upon BATSE measurements. This model was fit to the data for 
a grid of trial frequencies and numbers of Fourier coefficients $n$. The 
best-fit frequency was determined by the H-test (\cite{DeJager89}). Background
variations observed in these intervals were at low frequencies relative to the
Cep X--4's pulse frequency. Deviations from the constant background model
resulted in low frequency noise well outside the search range used which did
not affect pulse frequency measurements.
 
\subsection{Results}

\begin{figure}
\centerline{
\psfig{file=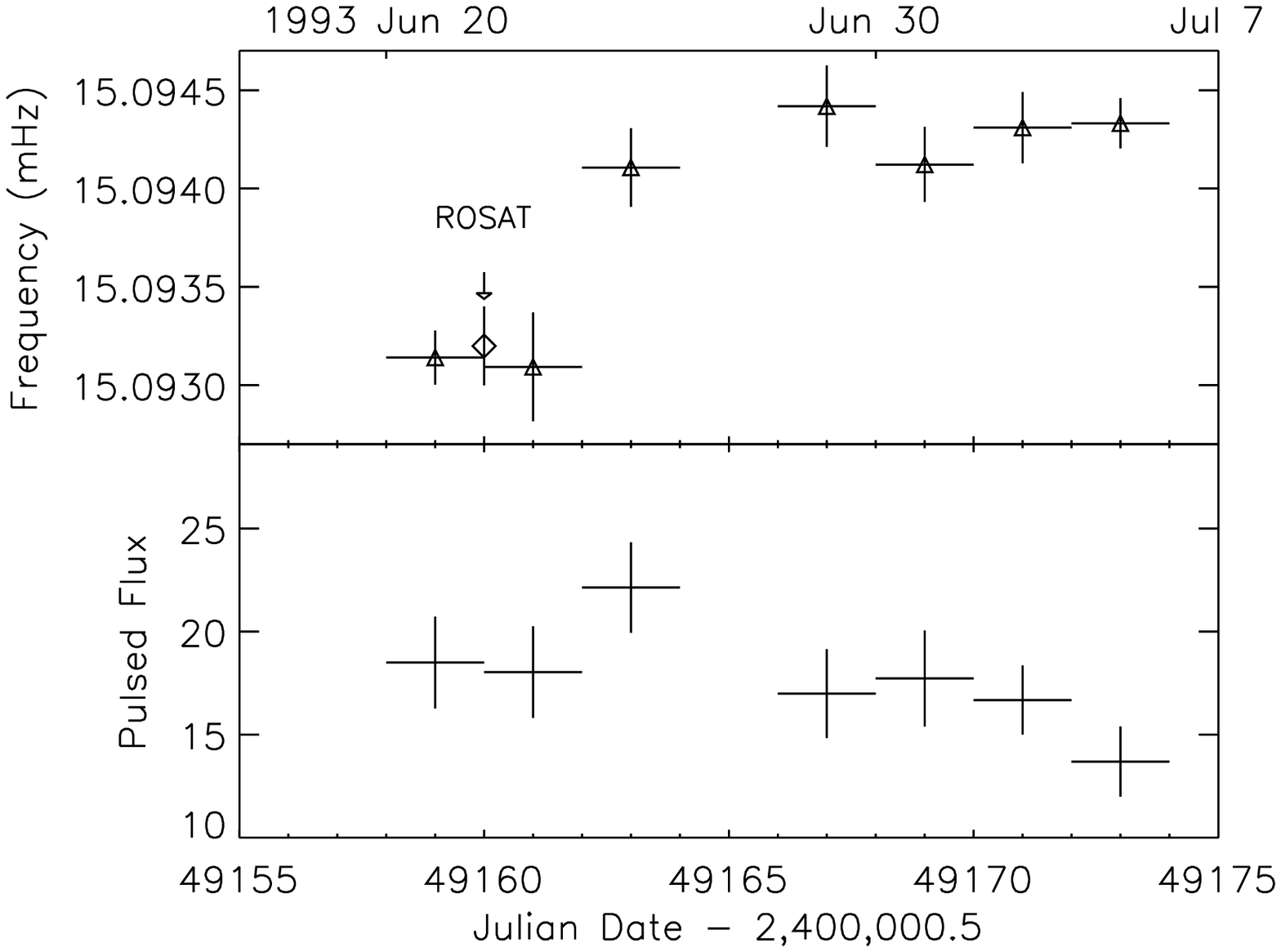,height=2in,width=3.5in}}
\caption{Cep X--4 pulse frequency and pulsed flux measurements determined at
2-day intervals using BATSE from 1993 June 20-July 6 (MJD 49,158-49,174). The 
pulse frequencies 
have been barycentered, but no binary correction has been applied, since the 
orbit is unknown. The square denotes the frequency measurement by ROSAT
(\protect\cite{Schulz95}). The rms pulsed fluxes (20-50 keV) in the center 
panel were calculated at 2-day intervals assuming an exponential spectrum 
(equation \protect\ref{eqn:spec}) with $E_{\rm fold} = 15$ keV (1 mcrab
=6.3$\times 10^{-3}$ keV cm$^{-2}$ s$^{-1}$). \label{fig:cepx4_93}}
\end{figure}

BATSE data from 1991 April to 1997 December were searched for outbursts from 
Cep X--4 using the pulse frequency and flux estimation method described in
section \ref{sec:tech}. Two outbursts, each lasting about 2 weeks were detected
in the BATSE data: the first from 1993 June 20 - July 6 (MJD 49,158-49,174) and
the second from 1997 July 3-15 (MJD 50,632-50,644). Figure~\ref{fig:cepx4_93} 
shows the history of pulse frequency and pulsed flux for the 1993 outburst. 
{\em ROSAT} also observed this outburst and measured a pulse frequency of 
$15.0931 \pm 0.0002$ Hz s$^{-1}$ (\cite{Schulz95}) which agrees well with
the BATSE measurements. From 
figure~\ref{fig:cepx4_93} one observes that Cep X--4's frequency increased
significantly from 1993 June 24-28 (MJD 49,162-49,166). The frequency rate is 
$\sim~10^{-12}$~Hz~s$^{-1}$, which may be intrinsic or due to orbital motion.

\begin{figure}
\centerline{
\psfig{file=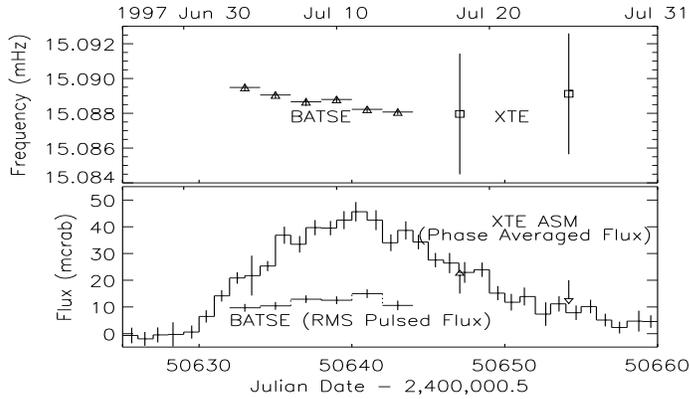,height=2in,width=3.5in}}
\caption{Pulse frequency and flux measurements from the 1997 July 1 - 1997
Aug 15 (MJD 50,630-50,657) outburst of Cep X--4. The pulse frequencies in the 
top panel, denoted by
diamonds, were determined at 2-day intervals from 20-50 keV BATSE DISCLA data.
These pulse frequencies have been barycentered, but no binary correction has 
been applied. The squares denote pulse frequency measurements from {\em RXTE} PCA 
data. The errors on the PCA frequency measurements have been corrected for
excess aperiodic noise. The solid lines in the bottom panel denote rms pulsed
fluxes (20-50 keV) measured at 2-day intervals with BATSE. The dotted lines 
are phase averaged 2-12 keV fluxes measured at 1-day intervals with {\em RXTE}
ASM (1 mcrab = 0.075 cts s$^{-1}$.) 
\label{fig:cepx4_97}}
\end{figure}

In 1997 July, Cep X--4 underwent another outburst which was detected
with BATSE and {\em RXTE}. Figure~\ref{fig:cepx4_97} shows the history of pulse
frequency as measured with BATSE and the {\em RXTE} PCA. The average frequency 
rate is $\sim -10^{-12}$ Hz s$^{-1}$, which is comparable in magnitude to the 
brief spin-up episode seen in the 1993 outburst.  This spin-down may be
intrinsic or due to orbital motion. Also shown is the pulsed flux history from
20-50 keV BATSE data and the phase 
averaged flux from {\em RXTE} ASM data. To determine the pulse frequencies from
the PCA data, Standard 1 (125 ms, no energy resolution) data from 1997 July 18
(MJD 50,647) and 1997 July 25 (MJD 50,654) were used with techniques described
in section~\ref{sec:tech}. We found that the scatter in frequency measurements
for different values of the number of Fourier coefficients, $n$ 
(see section~\ref{sec:tech}), was much larger than expected for Poisson noise.
The best fit frequencies, with errors corrected for aperiodic noise (described
below), were 15.088$\pm$0.004 mHz ($n=14$) on July 18 (MJD 50,647) and 
15.088$\pm$0.003 mHz ($n=14$) on July 25 (MJD 50,654). 

Excess aperiodic noise from Cep X--4 was likely the cause of the observed
scatter in our frequency measurements, so we performed power spectral analyses
on about 2000~s of Good Xenon data from 1997 July 18 and about 2400~s from 
July 25 to search for aperiodic variability. These data were binned into 125 
ms time bins containing no gaps. The background in these intervals
was modeled using pcabackest\footnote{\footnotesize In this paper for all background models, 
pcabackest version 1.5a was used with the Q6, Activation, and Cosmic X-ray 
models. For the fainter July 25 observation, pcabackest v2.0c was also tried 
with the L7 and 240 minute models for faint sources. Differences between the 
models were very small relative to source count rates, and did not significantly
affect any of our results. Details of these models are available from the {\em
RXTE} Guest Observer Facility.} , which estimates the background from Standard~2
data at 16 second intervals. This model was then interpolated to 125 ms
resolution by polynomial fitting. Power due to the 66~s pulsations was removed
by subtracting a model consisting of a 14th degree Fourier expansion in the 
best fit pulse frequency.  Next a Fourier transform with no oversampling was 
performed on the background subtracted data. The Fourier coefficients $a_k$
are given by
\begin{equation}
 a_k = \sum^N_{j=1} (C_j-B_j) e^{i 2 \pi j k/N}
\end{equation}
where $C_j$ and $B_j$ are the total counts and background counts in bin $j$,
$N =$ the number of measurement bins, and $i=(-1)^{1/2}$. Equation (1) in 
Morgan \& Remillard (1997) \nocite{Morgan97} was used to calculate the 
dead-time corrected Poisson level with a time bin size of $t_{\rm b} = 125$~ms,
a dead-time per event of $\tau = 10\ \mu$s, and a dead-time window per very
large event (VLE) of $\tau_{\rm vle} = 170\ \mu$s. In the normalization of 
Leahy et al. (1983) \nocite{Leahy83}, given by 
\begin{equation}
P_{\rm Leahy}= 2\frac{|a_k|^2}{N_{\rm ph}}
\end{equation}
where $N_{ph}$ is the total number of photons, the Poisson level is 
$P_{\rm Poisson}=2$ for purely Poisson noise. These calculations yielded 
average dead-time corrected Poisson levels of 
$P_{\rm Poisson} = 1.97$ for July 18 and $P_{\rm Poisson} = 1.96$ for July 25.
The aperiodic noise at the pulse frequency and its harmonics was 
estimated by averaging the source power, $P_{\rm Leahy}$, in the region 
$n\nu_o-\nu_o/2 \leq \nu \leq n\nu_o +\nu_o/2$, where $\nu_o$ is the pulse 
frequency (n=1) and $n\nu_o$ (n=2,3,...) are the harmonics of the pulse 
frequency. This power was then compared to the average dead-time corrected 
Poisson level, $\bar P_{\rm Poisson}$ for each region. The errors on the 
frequency measurements, $\sigma_{\nu}$,listed above and shown in 
figure~\ref{fig:cepx4_97} include a correction due to this excess aperiodic 
noise, given by $\sigma_{\nu} = (\bar P_{\rm Leahy}/\bar P_{\rm Poisson})^{1/2} 
\sigma_{\rm Poisson}$, where $\sigma_{\rm Poisson}$ is the error expected for 
Poisson noise only.  

\begin{figure}
\centerline{
\psfig{file=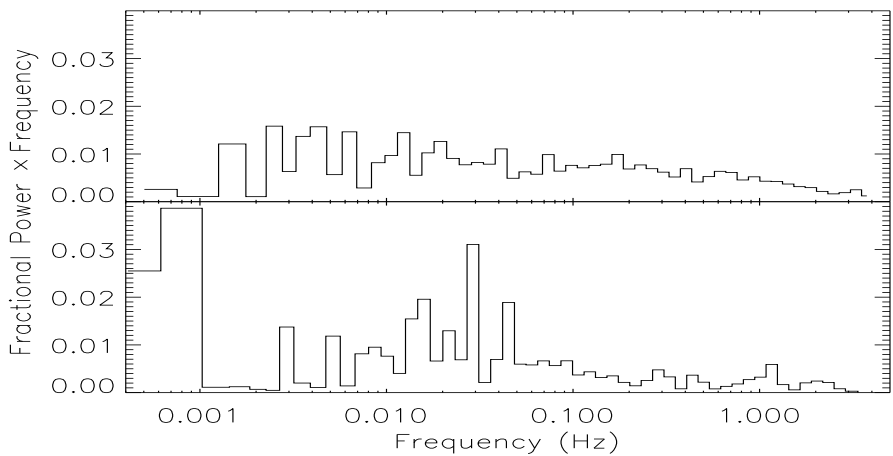,height=2.5in,width=5.0in}}
\caption{Power Spectra from 1997 July 18 (top panel) \& July 25 (bottom
panel) PCA observations. The mean pulse profile was subtracted from the data 
before calculation of the power spectrum.  The Poisson level has been
subtracted. The power is normalized to the mean source counts squared and has 
been multiplied by the analysis frequency to give the fractional variance per
logarithmic frequency interval.\label{fig:pow}}
\end{figure}

Next, we wished to measure the rms variation of the aperiodic noise component.
We renormalized the power such that the fractional rms variation in source 
intensity over a given frequency range is related to the integral of the power
in that range, i.e.
\begin{equation}
\frac{\sigma^2_S}{\bar S^2} = \int^{\nu_2}_{\nu_1} P(\nu) d\nu
\end{equation}
where $\sigma^2_S$ is the variance on the source count rate due to the
frequency range $[\nu_1,\nu_2]$ and 
$\bar S = \frac{1}{N}\sum^N_{j=1} (C_j-B_j)$ is the mean source count rate. 
The power spectrum was then calculated by
\begin{equation}
P_k = (P_{\rm Leahy} - P_{\rm Poisson}) \frac{N_{\rm ph}T}
{\left(\sum^N_{j=1} (C_j-B_j)\right)^2} 
\end{equation}
where $T$ is the duration of the data and $\sum^N_{j=1} (C_j-B_j)$ is the total
source counts in bin $j$. Figure~\ref{fig:pow} shows the power spectra, 
plotted as fractional power per logarithmic frequency interval, from 1997 
July 18 in the top panel and July 25 in the bottom panel. In these units, a
$1/f$ power law is a horizonal line. Aperiodic 
variability is present in both time intervals, with an rms variation of 17-19\%
($\nu=$0.01-1 Hz) for both days. In the 1988 outburst, {\em Ginga} observed 
aperiodic noise with an rms variation of about 6\% for the same frequency interval. The source was apparently noisier in 
{\em RXTE} observations. 

\begin{figure}
\centerline{
\psfig{file=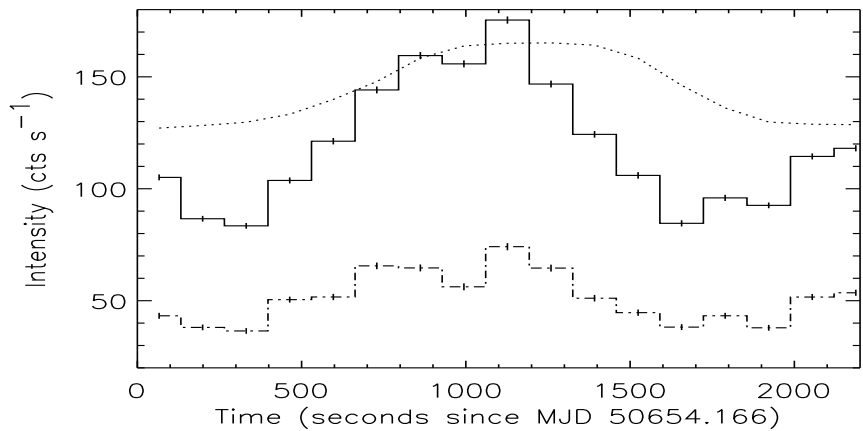,height=2.5in,width=5.0in}}
\caption{Comparison of source and background variations on 1997 July 25 (MJD
50,654). The background subtracted unpulsed flux (solid histogram), pulsed flux
(dotted histogram), and background model (dotted curve) were computed for
intervals exactly 2 pulse periods long. The pulse fraction is approximately
constant across these observations, indicating that the observed variations in
source flux are intrinsic to Cep X--4. \label{fig:2period}} 
\end{figure}

Clearly evident in figure~\ref{fig:pow} is the
presence of considerable low frequency noise on 1997 July 25 (bottom panel)
which is not present in the top panel. We compared variations in the pulsed
flux, mean background subtracted flux, and the background model (generated
using pcabackest) to determine if these features were intrinsic to Cep X--4. 
The unpulsed flux and pulsed flux for July 25 were measured by fitting a 
constant plus a 14 harmonic Fourier expansion in the best-fit pulse frequency 
to intervals exactly 2 pulse periods long. The background model was binned at
the same resolution and subtracted from the unpulsed flux. The variations 
observed in the unpulsed flux prior to background subtraction were about 3-4 
times the background model variations. Figure~\ref{fig:2period} shows the 
background model, the background subtracted unpulsed flux, and the pulsed flux.
The background subtracted unpulsed flux 
(denoted by the solid histogram) and pulsed flux (denoted by the dashed
histogram) increased slowly to a broad peak, followed by a faster decrease. The
background model (denoted by the dotted line) also slowly rose to a flat peak 
and then decreased, but it peaked later. The pulse fraction remained 
essentially constant throughout this interval, so it was unlikely that the 
variations observed were due to background fluctuations.  On July 18, when Cep
X--4 was about 2 times brighter in the ASM, the unpulsed and pulsed flux
remained approximately constant during our observations. Apparently Cep X--4 was
less stable at lower intensities.

\begin{figure}
\centerline{
\psfig{file=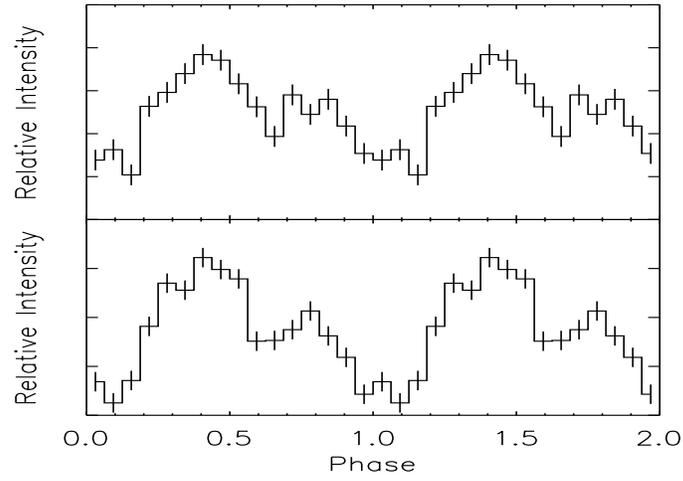,height=2.5in,width=5.0in}}
\caption{Average 20-50 keV pulse profiles from (top panel) 1993 June 20-29 
(MJD 49,158-49,167) and (bottom panel) 1997 July 4-14 (MJD 50,633-50,643)
BATSE data.\label{fig:batse_profiles}}
\end{figure}

Average pulse profiles were generated by epoch-folding $\sim$ 10 days of 20-50
keV BATSE CONT (2.048 s resolution data) for both the 1993 and 1997 outbursts.
A quadratic frequency model fit to the observed frequencies in 
Figures~\ref{fig:cepx4_93}~\&~\ref{fig:cepx4_97} was used in the epoch-folding.
Pulse profiles from 1993 June 20-29 (MJD 49,158-49,167) and 1997 July 4-14 (MJD
50,633-50,643) are shown in the top and bottom panels respectively in
Figure~\ref{fig:batse_profiles}.  Both profiles show a large main peak (phase
$\sim 0.2-0.6$) followed by smaller second peak atop a broad shoulder (phase
$\sim 0.6-0.9$) and a valley (phase $\sim 0.9-1.2$).

Pulse profiles from {\em RXTE} observations were generated by epoch-folding 
Good Xenon data from 1997 July 18 and July 25. Good Xenon data were binned into
125 ms bins and divided into 8 energy bands. The background in each band was 
estimated at 16 s intervals using pcabackest and interpolated to 125 ms 
resolution by polynomial fits. The background subtracted data were folded at 
the best fit frequencies from the fits to the Standard 1 data shown in 
figure~\protect\ref{fig:cepx4_97}. Figure~\ref{fig:profiles} shows the pulse 
profiles from July 18 (left hand panel) and July 25 (right hand panel). The 
pulse profile shows energy dependent features and considerable structure with 
a large main peak (phase $\sim$0.7-1.0) followed by a deep notch at low 
energies (phase $\sim$1.0-1.1). The notch is followed by a second peak 
(phase $\sim$1.1-1.4) and a valley (phase $\sim$1.4-1.7). The profiles evolve 
from double peaked, with a narrow notch between the peaks at low energy, to a 
dominant first peak (phase $\sim 0.7-1.0$) followed by a broad shoulder. The 
second peak (phase $\sim 1.1-1.4$) appears to be weaker relative to the first 
peak on July 25 than on July 18. 

\begin{figure}
\centerline{
\psfig{file=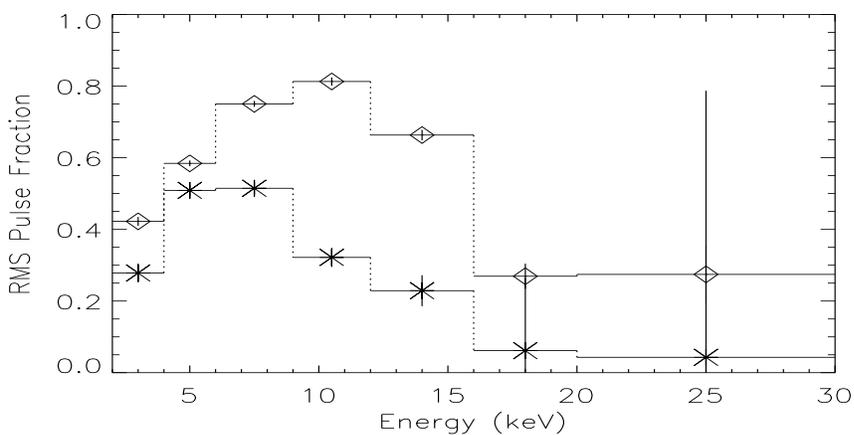,height=2.5in,width=5.0in}}
\caption{The root-mean-squared (RMS) pulse fraction from {\em RXTE} PCA
observations on 1997 July 18 (diamonds) \& 25 (asterisks). The 
pulse fraction is energy dependent and appears to be decreasing with intensity.
\label{fig:frac}} 
\end{figure}

Figure~\ref{fig:hardness} is a hardness ratio for July 18 
given by
\begin{equation} 
r=\frac{h+\Delta h}{s +\Delta s}
\end{equation}
where $h$ is the average background subtracted count rate for 9-20 keV, $\Delta 
h$ is the mean subtracted pulse profile count rate for 9-20 keV, $s$ is the 
average background subtracted count rate for 2-6 keV, and $\Delta s$ is the
mean subtracted profile for 2-6 keV.  The errors shown are the errors on the
difference $r-h/s$, since we are interested in where the hardness ratio
deviated from the mean value, 0.584$\pm$0.002. Hardness ratios show that the 
notch is the hardest feature and that it is surrounded by a softer double lobed 
feature (phase $\sim$ 0.7-1.4). Similar features were also seen in hardness
ratios from July 25. However, the significance of these features was reduced by
the lower intensity and smaller pulse fractions in these data.
The root mean squared (RMS) intensities
for each energy band were calculated as $s_{\rm rms} = (\sum_{j=1}^N (s_j-\bar
s)/N)^{1/2}$, where $s_j$ is the background subtracted source count rate in 
phase bin $j$, $N=32$ is the number of phase bins, and $\bar s$ is the mean 
source count rate. RMS pulse fractions, $s_{\rm rms}/\bar s$ are shown in 
figure~\ref{fig:frac}. The pulse fraction has considerable energy dependence 
and appears to be decreasing as the mean source intensity decreases.  

The average 20-50 keV BATSE pulse profile from 1997 July 4-14
(Fig.~\ref{fig:batse_profiles}, bottom panel) is very similar 
to the 1997 July 18 PCA profile in the 20-30 keV energy band 
(Fig.~\ref{fig:profiles} bottom left). The source intensity was too low to 
allow detection of significant profile variations at energies $>$ 20 keV with
either instrument. {\em Ginga} observed pulse shape variations with energy 
during the 1988 outburst. The second peak was brighter than the first at low 
energies (1-7 keV) and roughly equal at higher 
energies (13-37 keV) (\cite{Koyama91}).   

\section{Discussion}

The optically identified companions for transient pulsars with periods longer
than one second are all Be or Oe stars. Recent long-term studies with BATSE
revealed that Be/X-ray binaries exhibit series of often periodic outbursts and
most appear to have a long-term spin-up trend (\cite{Bildsten97}).  One other
system, A 1118-616, may be experiencing a spin-down trend, but the duty cycle 
for this source, like Cep X--4, is quite low. Long-term spin-down in X-ray
pulsars is thought to be caused by the propeller effect (\cite{Illarionov75}).
Simple accretion theory assumes that material from the companion star is
flowing onto a rotating neutron star with a magnetic field. The magnetic field
is so strong that it determines the motion of the material in a region of space
surrounding the neutron star called the magnetosphere. 
The magnetospheric radius $r_{\rm m}$, is defined to be the distance from the 
neutron star at which all magnetic field lines are closed loops. Assuming a
dipole field at large distances from the neutron star, the magnetospheric
radius can be estimated by equating the magnetic energy density to the total
kinetic energy density of the accreting gas. Therefore from Shapiro \&
Teukolsky (1983) \nocite{Shapiro83},
\begin{equation}
  r_{\rm m} = 3.09 \times 10^8\, {\rm cm}\, \epsilon \mu_{30}^{4/7} m^{-1/7} \dot
M_{17}^{-2/7} 
\end{equation}
where $\mu_{30}$ is the magnetic moment of the dipole field in units of
$10^{30}$ G cm$^3$, $m$ is the neutron star mass in units of $1.4 M_{\sun}$,
$\dot M_{17}$ is the mass accretion rate in units of $10^{17}$ g s$^{-1}$, and
$0.5 \lesssim \epsilon \lesssim 1$ for disk accretion (\cite{Ghosh79}; 
\cite{Arons93}; \cite{Ostriker95}; \cite{Wang96}) and $\epsilon = 1$ for wind 
accretion (\cite{Shapiro83}).
When $r_{\rm m}$ exceeds the corotation radius, $r_{\rm co}$, the distance 
from the neutron star where centrifugal forces just balance local gravity,  
the propeller effect is believed to occur. Matter will be ejected from the 
magnetic field lines when the centrifugal forces exceed gravitational force. 
This ejected matter gains angular momentum from the pulsar, causing it to spin
down. If we assume the long-term spin-down observed in this system is due to 
the propeller effect, then by equating the magnetospheric and co-rotation 
radii, we can estimate the critical mass accretion rate where the propeller
effect begins 
\begin{equation}
\dot M_{\rm crit} = 8.41 \times 10^{17}\, {\rm g\,s}^{-1}\, \epsilon^{7/2}
\mu_{30}^2 m^{-5/3} P^{-7/3}
\end{equation}
where $P$ is the pulsar spin period in seconds.  At this critical mass
transfer rate, Corbet (1996) predicts a transition from neutron star
accretion to magnetospheric accretion, accompanied by a large drop in
luminosity. Using a magnetic field estimate of $B = 2.6 \times 10^{12}$ G 
(\cite{Mihara91}), a distance estimate $d = 3.8 \pm 0.6$ kpc 
(\cite{Bonnet98}), a spin period of $P = 66.25$ s, along with typical neutron 
star parameters $m = 1$ and $R = 10^6$ cm, we estimate the minimum X-ray flux 
for neutron star accretion, $F_{\rm ns, min} = 9.1 \times 10^{-12} \epsilon^{7/2}$ erg cm$^{-2}$ s$^{-1}$,   
and the maximum for magnetospheric accretion, $F_{\rm m, max} = 3.3 \times 10^{-15} 
\epsilon^{7/2}$ erg cm$^{-2}$ s$^{-1}$. In 1993 January {\em ROSAT} observed
a quiescent 0.1-2.5 keV flux level of $2.8 \times 10^{-13}$ erg cm$^{-2}$ s$^{-1}$ from Cep
X--4 (\cite{Schulz95}). This flux level fell in the gap between $F_{\rm ns, min}$ and $F_{\rm m,
max}$. Schulz et al. (1995) \nocite{Schulz95} estimated that the {\em ROSAT}
flux in the outburst in 1993 June was about 10\% of the bolometric flux.
Assuming that the Cep X--4 spectrum was unchanged during the quiescent 
{\em ROSAT} observations, we estimate a quiescent bolometric flux level of 
$2.8 \times 10^{-12}$ erg cm$^{-2}$ s$^{-1}$, which still falls in the gap, if
we assume wind accretion
($\epsilon = 1$). Clearly the observed flux was near the expected value for the
onset of the propeller regime. The estimated bolometric flux from the {\em ROSAT}
observations was a factor of $\sim$ 3 below the expected value for wind
accretion, which implies that the spectrum was much harder at low $\dot M$,
similar to A 0535+262 (\cite{Giovannelli92}). Future quiescent observations
at higher energies are needed to confirm this spectral hardening. For wind
accretion, additional uncertainties in the predicted flux arise because when
the propeller effect sets in at the neutron star's equator, accretion still
occurs at higher latitudes. Comparing the estimated bolometric flux with the
predicted flux suggests that we may be observing disk accretion. However,
additional uncertainties in the predicted flux arise from uncertainty in
dimensionless parameter $\epsilon$. In both wind and disk accretion, the
validity of the assumptions used to derive the predicted flux is uncertain
because the physics at the onset of the propeller effect is poorly understood.

Long-term studies show that Be/X-ray transients exhibit two types of outbursts,
giant outbursts characterized by high luminosities and high spin-up rates, and
normal outbursts which typically are fainter, exhibit less intrinsic spin-up, 
and are spaced by some multiple of the orbital period. Normal outbursts appear
to be associated with peristron passage in Be/X-ray transients with known 
orbits (\cite{Bildsten97}). The outbursts observed from Cep X--4 are most likely
normal outbursts. Therefore the spacing of the 4 outbursts constrains the
orbital period. We assume that the two previous outbursts in 1972 June-July 
(\cite{Ulmer73}) and 1988 March-April (\cite{Koyama91}) occurred at the same 
orbital phase (likely near periastron) as the two outbursts observed by BATSE.
Outburst mid-times were calculated for each observation. Errors were assigned
to each mid-time according to how well the mid-point of the outburst was known.
An evenly spaced grid of orbital periods ranging from 25-10$^6$ days was searched, minimizing
the following $\chi^2$:            
\begin{equation}
\chi^2 = \sum_{k=1}^4 \frac{|1 - e^{i2 \pi (e_k - e_0)/ P_{\rm orb}}|^2}{(2 \pi
\sigma_k/ P_{\rm orb})^2}.
\end{equation}
where $e_k$ is the mid-time of outburst $k$, $\sigma_k$ is the estimated error
on $e_k$, $e_0$ is an epoch, and $P_{\rm orb}$ is an orbital period. A number
of acceptable orbital periods, all $\lesssim$ 147.3 days, were found.
Combining this result with Koyama et al. (1991) \nocite{Koyama91}, implies 23
days $\lesssim P_{\rm orb} \lesssim 147.3$ days.  
                                                        
\begin{figure}
\centerline{
\psfig{file=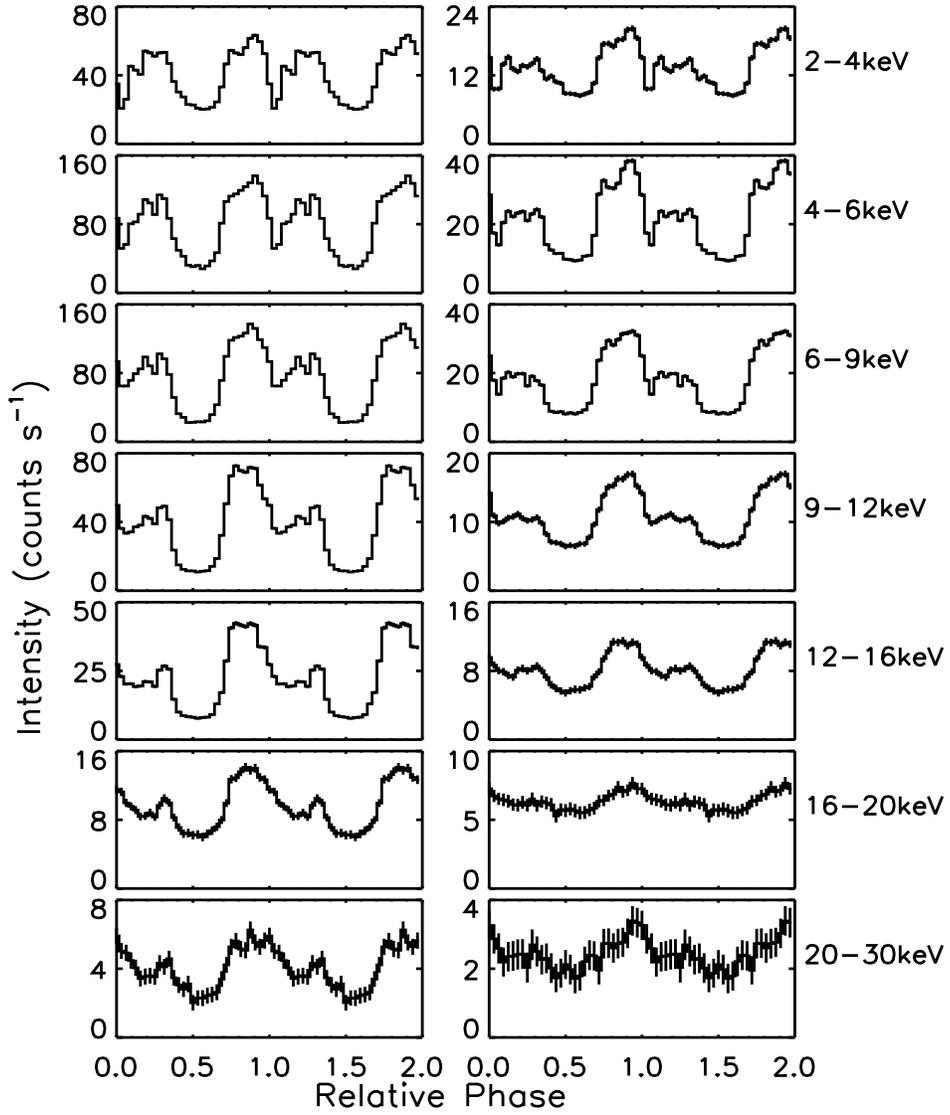,height=6in,width=5.0in}}
\caption{Pulse profiles from the {\em RXTE} PCA observations on 
1997 July 18 \& 25 (MJD 50,647 \& 50,654).\label{fig:profiles}} 
\end{figure}

\begin{figure}
\centerline{
\psfig{file=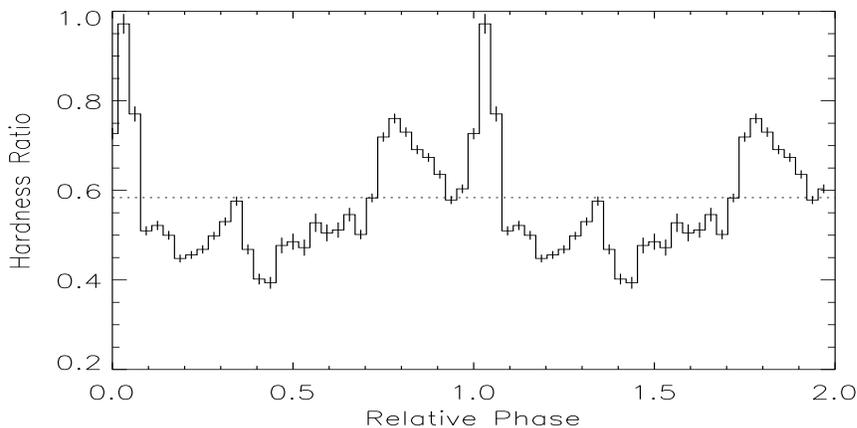,height=2.5in,width=5.0in}}
\caption{The hardness ratio between 9-20 keV background subtracted 
count rates and 2-6 keV background subtracted count rates on July 18 (MJD
50,647). The 
errors shown are errors on the difference between this ratio and the average 
hardness ratio. The dotted line corresponds to the average hardness ratio of 
0.584$\pm$0.002. \label{fig:hardness}}
\end{figure}

The energy dependent pulse profile of Cep X--4 in figure~\ref{fig:profiles} 
consists of two peaks separated by a narrow valley at low energies ($<6$ keV) 
which becomes filled in as the energy increases, until the leading peak 
(phase $\sim$ 0.7-1.0) becomes dominant at energies above roughly 9 keV. This 
energy dependent behavior can be interpreted as the sum of two pulse 
components: a softer double lobed component that occupies pulse phase $\sim$ 
0.7 to 1.4 in figure~\ref{fig:profiles} and a 
narrower harder peak centered at phase $\sim$ 1.05 that is offset towards the 
leading peak of the softer double lobed component. The relationship between the
two components is well illustrated by the hardness ratio shown in
figure~\ref{fig:hardness}. This energy dependent behavior is quite similar to 
that observed in Her X-1 (\cite{Deeter98}). 

A  possible interpretation of this energy dependent behavior can be made using 
a cyclotron backscattering scenario. In this scenario the outward bound X-rays 
from a given magnetic pole are divided into two main components: a hard pencil
beam directed along the accretion column and an antipodally directed fan beam 
of backscattered radiation (\cite{Brainerd91}). The fan beam is 
produced by outward bound photons that travel up through the accretion column 
until their energy equals the local cyclotron resonance energy whence they are 
backscattered. The backscattered photons will either strike the neutron
star surface or be focused around the neutron star forming an antipodal fan beam.
Softer photons scatter from higher altitudes in the accretion column and hence 
a smaller fraction of the backscattered photons are intercepted by the neutron
star surface than is the case for the harder photons that scatter at lower
altitudes. Therefore, the antipodal fan beam lobes produced by the softer 
photons will be wider, i.e. more ``filled in" than the lobes produced by harder
photons.  At even higher energies, near the surface cyclotron 
energy and above, most photons will not be backscattered and will escape 
forming a pencil beam. For an appropriate angle of the magnetic axis to the 
spin axis and an observer positioned with a line-of-sight to the neutron star 
near the spin latitude of the magnetic dipole axis, a pulse profile consisting
of a single hard pencil beam filling in the center of a softer fan beam from
the antipodal pole will be observed.    

\begin{figure}
\centerline{
\psfig{file=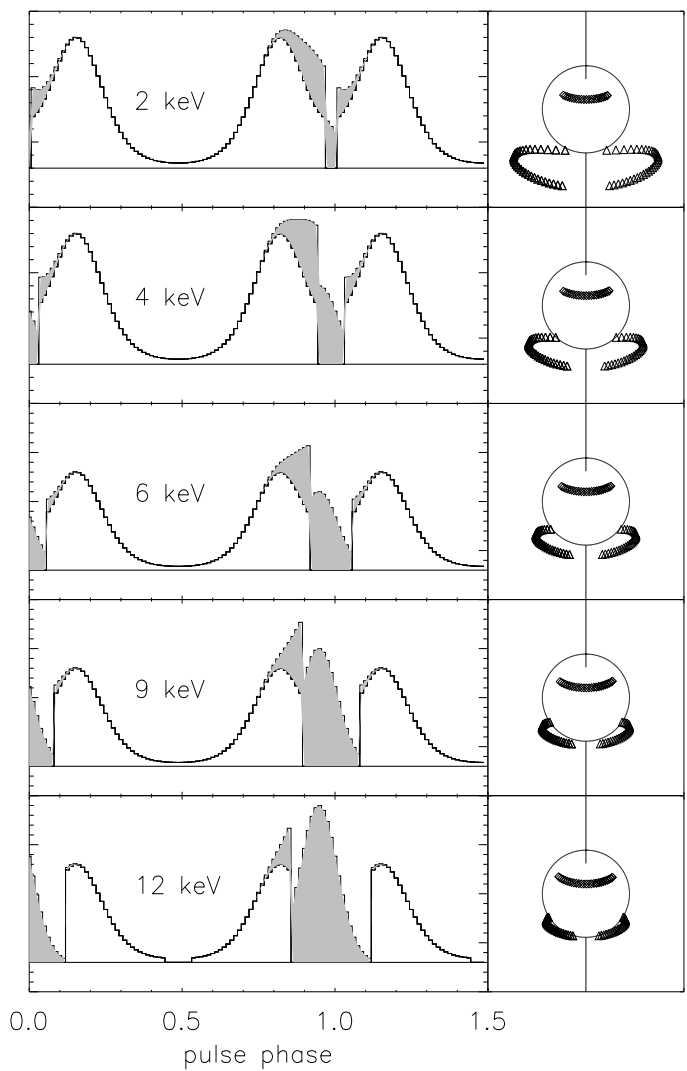,height=6in,width=6.0in}}
\caption{
(Left panels) Pulse profiles produced by a simple cyclotron scattering beam
model for the five photon energies 2, 4, 6, 9 and 12 keV (from top to bottom).
The pencil beam contributes the gray shaded area to the pulse profile while
the fan beam contributes a double humped profile. Artificially sharp drops in
the profile (e.~g. phase $\sim$ 1.1) result from simplified modeling of
emission as from a point rather than an extended region. (Right panels) Visible 
locations of the points emitting the pencil and fan beam profiles. The neutron
star is the circle with the spin axis shown by the vertical line. The pencil
beam point is on the surface of the neutron star with the fan beam point 
located above the surface. The angle between the observer and spin axis is
$61^{\circ}$ and magnetic axis and spin axis is
$28^{\circ}$.\label{fig:pulsemod}} 
\end{figure}

A simple beam model consisting of a pencil beam surrounded by a backscattered
fan beam was constructed to explore possible pulse profiles illustrative of
a cyclotron scattering model. The pencil beam is represented by an emitting
point on
the neutron star surface and has a gaussian beam profile. Similarly, the fan
beam is represented by an emitting point above the neutron star surface that
is at the apex of a conical beam pointed toward the neutron star and has a
gaussian profile for the conical beam ``walls". Gravitational light bending
effects are taken into account by calculating photon trajectories from that
height. In figure~\ref{fig:pulsemod} we show an example profile that is reminiscent of the
Cep X--4 profile. The five panels on the left represent the pulse profile in
five different energy bands. The panels on the right show the apparent
locations of the emitting point when the beams are visible.
The height $R$ of the fan beam emitting point decreases with increasing
energy as $E_c (r) = 11.6 B_s \left(\frac{R_s}{R}\right)^3$ keV where $E_c$ is the
cyclotron scattering energy, $B_s$ is the surface magnetic field strength in
units of $10^{12}$ Gauss and $R_s$ is the radius of the neutron star.
Sharp drops in the fan beam emission occur when the beam intersects the
neutron star surface. Use of a more realistic extended emission region rather
than a point would smooth out these sharp drops.The pencil beam increases in strength relative to the
fan beam with increasing energy and contributes the shaded area to the pulse
profile.
The observer is positioned such that only a single pencil beam is visible
from one magnetic pole while only the fan beam from the opposite pole is
visible.
The pole emitting the pencil beam has been offset
slightly to move the pulse phase centroid of the resulting pulse profile
with respect to the pulse profile contributed by the fan beam. This is done
to simulate the asymmetry observed in the Cep X--4 profile.

\acknowledgements{This research has made use of public target of opportunity
PCA data obtained from the {\em RXTE} GOF. The quick-look ASM data on Cep X--4
were provided by the {\em RXTE}/ASM team at MIT and NASA/Goddard Space Flight 
Center.}

\newpage

\end{document}